\documentstyle[12pt]{article}
\newcommand{\be}{\begin{equation}}
\newcommand{\ee}{\end{equation}}
\newcommand{\ba}{\begin{eqnarray}}
\newcommand{\ea}{\end{eqnarray}}
\newcommand{\baa}{\begin{eqnarray*}}
\newcommand{\eaa}{\end{eqnarray*}}
\newcommand{\bb}{\begin{thebibliography}}
\newcommand{\eb}{\end{thebibliography}}
\newcommand{\ci}[1]{\cite{#1}}
\newcommand{\bi}[1]{\bibitem{#1}}

\begin{document}
\phantom{.}
\vspace{2cm}

\flushright { {\large Paper:JINR-E2-93-433}}
\vspace{2cm}
\begin{center}
{ \large THE BEHAVIOR OF THE SLOPE OF ELASTIC NUCLEON \\
        SCATTERING  AT SMALL TRANSFER MOMENTA \\
         AND RECENT UA4 DATA      } \\
\vspace{2cm}
S.V.Goloskokov, S.P. Kuleshov, O.V.Selyugin\\
Bogolyubov Laboratory of Theoretical Physics,\\
Joint Institute for Nuclear Research, Dubna\\
Head Post Office P.O.Box 79, 101000 Moscow, Russia \\
\end{center}

\vspace{2cm}

\begin{abstract}
 Theoretical predictions for the behavior of the slope of the nucleon-nucleon
scattering and another parameters of the differential cross sections
in the framework of the dynamic model
are compared with the recent UA4 data at small transfer momenta
and at the centre-of-mass energy of 541 GeV.
The predictions at superhigh energies are considered.
\end{abstract}

\newpage
\phantom{.}
\vspace{1cm}
      An actual problem of the modern physics of elementary
particles,  the research of strong interaction processes at large
distances and  high energies, is considered in the framework of different
approaches by using various models of the structure of hadrons and the
dynamics of their interactions (see e.g. \ci{bas}).
	The diffraction scattering cannot yet be described quantitatively
in the framework  of the perturbative QCD. Therefore, it necessary
to apply the different models  which can describe the hadron-hadron
interaction at large distances \ci{sjg}.
The research  of elastic  scattering requires the knowledge
of properties of the pomeron,
the object determining  the interaction of hadrons in this  range.
In this case  the study of the structure  and  spin  properties of
both the hadron  and the pomeron  acquires  a  special  role.
In a number of works [3-5] the dynamical model with taking account of
the interaction at large distances has been developed.
The model is based on the general quantum field theory principles
(analyticity, unitarity and so on) and takes into account basic information
on the structure of a hadron as a compound system with a central part region
where the valence quarks are concentrated and a long-distance region
where the color-singlet quark-gluon field occurs.

Let us consider the nucleon-nucleon scattering.  In  \ci{g2}  on
the basis of sum rules it has been shown that the main contribution  to
hadron interaction at large  distances  comes  from  the  triangle
diagram with $2\pi $ -meson exchange in the $t$-channel. As a result, the
hadron amplitude can be  represented  as  a  sum  of  central  and
peripheral parts of the interaction:
\ba
T(s,t) \propto T_{c}(s,t) + T_{p}(s,t)
\ea
 where $T_{c}(s,t)$   describes   the
interaction between the central parts of hadrons; and   $T_{p}(s,t)$  is
the sum of contributions of diagrams corresponding to
the interactions of the central part of one hadron  on  the  meson
cloud of the other. The contribution of these diagrams
to the scattering amplitude  with  an $N(\Delta $-isobar)  in  the
intermediate state looks as follows \ci{zpc}:

\ba
T ^{\lambda _{1}\lambda _{2}}_{N(\Delta )}(s,t) =
{g^{2}_{\pi NN(\Delta )}\over i(2\pi )^{4}}\int
 d^{4}q T_{\pi N}(s\acute{,}t)\varphi _{N(\Delta )}
[(k-q),q^{2}]\varphi _{N(\Delta )}[(p-q),q^{2}]  \nonumber \\
\times \frac{\Gamma^{\lambda_{1}\lambda_{2}}(q,p,k,)}
{[q^{2}- M^{2}_{N(\Delta )}+ i\epsilon ][(k-q)^{2}- \mu ^{2}+i\epsilon ]
[(p-q)^{2}- \mu^{2}+ i\epsilon ]}.
\ea

Here $\lambda_{1},\lambda_{2}$  are  helicities  of  nucleons;
$T_{\pi N}$  is  the
$\pi N$-scattering amplitude; $\Gamma$  is a matrix element of the numerator
of the representation of the diagram ; $\varphi $ are vertex functions chosen
 in the dipole form  with
the parameters $\beta _{N(\Delta )}$:

\ba
\varphi _{N(\Delta )}(l^{2},q^{2}\propto  M^{2}_{N(\Delta )}) = {\beta
^{4}_{N(\Delta )}\over (\beta ^{2}_{N(\Delta )}- l^{2})^{2}}.
\ea

        The model with  the $N $  and  $\Delta $
contribution  provides  a self-consistent  picture  of
the differential cross sections and spin phenomena
of different hadron processes  at  high  energies.
Really, parameters in the  amplitude  determined from one
reaction, for example, elastic $pp$-scattering, allow one to  obtain
a wide range of results for elastic
meson-nucleon scattering and charge-exchange reaction
 $\pi^{-} p \rightarrow  \pi^{0} n$
 at high energies.

 It is essential that  the
model predicts  large  polarization  effects  for  all  considered
reactions at high and superhigh energies \ci{zpc}. The predictions  are  in
good agreement with the experimental data  in  the  energy  region
available for experiment. Also note that just the effect of  large
distances determines a large value of the  spin-flip  amplitude  of
the charge-exchange reaction \ci{g4}.

  The model takes into account the $ s \rightarrow u$ crossing diagrams
in the scattering amplitude, which leads to the asymptotic equality of the
proton-proton and proton-antiproton cross sections as $s \rightarrow \infty$.
An important property of this model is that it can be applied
to the proton-antiproton scattering  at sufficiently low energies.
Thus,  the  behavior  of  the  proton-proton  and  proton-
antiproton differential cross sections at
 $p_{L}= 40 GeV$ and  $p_{L}= 1850 GeV$
 acquires a natural explanation \ci{g3}.

 The model gives the universal  behavior  of  the
diffraction peak slope for all hadron reactions \ci{g2}.
 More than a decade, the description  of the proton-proton
scattering at small transfer momenta and energies of ISR showed
the possibility
of oscillations  of the differential cross sections with a small
amplitude about several percent \ci{prot82}. A more careful analysis
of the differential cross sections in the coulomb region of transfer momenta
  also revealed this possibility \ci{sel}.
At the same time,
the model gave the understanding of the effect of oscillation revealed
in \ci{antip}. Such big oscillations are caused by changing the
slope of the differential cross sections with transfer momenta, on one hand,
and its description in terms of the exponential form,
on the other hand \ci{prot82}.
     The predictions of the model in the range of the second
diffraction peak coincide with the experimental data at $\sqrt{s} = 630 GeV$.
\ci{g1}

     Now let us compare our model predictions \ci{mars,g1,zpc}
with the recent data of UA4 collaboration \ci{ua4}.
The data are very precise and give errors only of several percent.
 The model predictions for the behavior of the slope of the
diffraction peak made for $\sqrt{s} = 540 GeV $ give
   $ B(s,t) = 15.5 Gev^{-2}$  at   $0<|t|<.15GeV^{2}$
and $\rho(s,t) = Im T(s,t)/Re T(s,t) = .14 $ at $|t|=.001Gev^{2}$.

As can be seen from fig.1, the model predictions are in
 good agreement with
 the new experimental data. We obtain $\chi^{2}/(2N) = 158/99$ with no
change of model parameters. However, as the value of the cross sections
in the model was normalized to the experimental data
in the range of energies of ISR which have the bias errors about 10 -20
percent,
we can insert a further norm about 5 percent
  into our model predictions.
Then,  $\chi^{2}/(2N) = 102/99$.
It should be noted that in the range of small transfer momenta the model
leads to a smooth change of the slope of the differential cross sections
 with growing $|t|$. For the energy range  under discussion
we have the following value of the slope (see Table I).

    Our predictions for the range of small transfer momenta
at $\sqrt{s}=1.8 TeV$
are shown in fig. 1.
 The values of the parameters
$\rho(s,t)$ and $ B(s,t)$
 for that energy are also presented
in Table I.
 As is obvious from Table I, both $\rho(s,t)$ and
$B(s,t)$  depend heavily on $s$ and $t$.

\vspace{.5cm}

{\bf TABLE I}
\vspace{1.5cm}

\begin{tabular}{|c|c|c|c|c|} \hline
{\em $ -t $} &
\multicolumn{2}{c|}{$ \sqrt{s} = 541 GeV $} & \multicolumn{2}{c|}{$ \sqrt{s} =
1800 GeV $} \\ \cline{2-5}
$ GeV^{2}$ & \  $ \rho( s,t)  $ \  & $B(s,t) GeV^{-2}$  & $ \; \rho (s,t) \; $
& $ B(s,t) GeV^{-2}$ \\ \hline
.001  &   .141      &    16.8    &    .182    &    18.1 \\
.014  &   .135      &    16.5    &    .178    &    17.7  \\
.066  &   .112      &    15.5    &    .161    &    16.6  \\
.120  &   .089      &    14.9    &    .143    &    15.9   \\ \hline
\end{tabular}
\vspace{1.5cm}

  At this energy we can see some oscillations of the cross sections with
a very small amplitude and the period depending on transfer momenta.
Now we can only remark that the possibility of existence of such
oscillations requires more
careful theoretical and experimental researches.

       The results which rather well describe the experimental data
at $\sqrt{s} = 540 GeV$ were obtained in the model \ci{sof}.
Note that like in
all modifications of the Chow-Yang model, it predicts the appearance
of a strongly marked diffraction structure at superhigh energies.
We must emphasize  that only our model gives such a large growth
both of differential and total cross sections and have no sharp dips
in the cross sections at superhigh energies.

Thus, it is clear that the most distinctive predictions of different models
 belong to the range of $|t| \sim 1. Gev^{2}$. So, our prediction
for the cross section at $\sqrt{s}=2.TeV$  and $|t|=1. GeV^{2}$
 $d\sigma / dt = 0.0213 mb/GeV^{-2}$ \ci{g2} and the prediction of
the model \ci{sof} $d\sigma / dt = 0.0112 mb/GeV^{-2}$ .
The difference of model predictions in the range of small transfer momenta
 is basically in the values of  $\rho(s,t)$ and its dependence on $s$ and
$t$.
It should be noted that the change of the parameters of
the differential cross sections at small transfer momenta
requires a very careful analysis of the procedure of determining
these parameters, especially in obtaining the value of
the total cross sections \ci{sel} as well as further theoretical
reserch and  measurement of the differential cross sections
in the coulombic range of superhigh energies.
\vspace{1cm}
     {\it Acknowledgement.}  {\hspace {0.5cm}}   The authors express their deep
gratitude
  to V.A. Matveev, V.A. Meshcheryakov,
   A.N. Sissakian
  for the support in this work; one of the authors
  (O.V. Selyugin) expreses his deep gratitude to L. Jenkovsky for
   fruitful discussions of the problems considered in this paper.

\newpage
\vspace{2cm}
\begin{center}
{\large                         Figure captions     } \\
\end{center}
 Fig. 1 -------------------- the model predictions for $\bar p p$ -
scattering.\\
\hspace{2cm}	     $\circ\!\!\!\mid$ - the experimental unnormalized data UA4
\ci{ua4}   \\
\hspace{2cm}	     - - - - - - - - - -  our predictions at  $\sqrt{s}=1.8 TeV$
 \end{document}